# MULTIFERROICITY IN THE MOTT INSULATING CHARGE-TRANSFER SALT κ-(BEDT-TTF)$_2$Cu[N(CN)$_2$]Cl

M. Lang[(a*)], P. Lunkenheimer[(b)], J. Müller[(a)], A. Loidl[(b)], B. Hartmann[(a)], N.H. Hoang[(a)], E. Gati[(a)], H. Schubert[(a)], and J. A. Schlueter[(c)]

[(a)] *Institute of Physics, Goethe-University Frankfurt(M), Germany*
[(b)]*Experimental Physics V, Center for Electronic Correlations and Magnetism, University of Augsburg, Germany*
[(c)]*Materials Science Division, Argonne National Laboratory, Argonne, Illinois 60439, USA*

**The recently proposed multiferroic state of the charge-transfer salt κ-(BEDT-TTF)$_2$Cu[N(CN)$_2$]Cl [P. Lunkenheimer *et al.*, *Nature Mater.*, vol. 11, pp. 755-758, Sept. 2012] has been studied by dc-conductivity, magnetic susceptibility and measurements of the dielectric constant on various, differently prepared single crystals. In the majority of crystals we confirm the existence of an order-disorder-type ferroelectric state which coincides with antiferromagnetic order. This phenomenology rules out scenarios which consider an inhomogeneous, short-range-ordered ferroelectric state. Measurements of the dielectric constant and the magnetic susceptibility on the same crystals reveal that both transitions lie very close to each other or even collapse, indicating that both types of order are intimately coupled to each other. We address issues of the frequency dependence of the dielectric constant ε' and the dielectric loss ε'' and discuss sample-to-sample variations.**

## I. INTRODUCTION

ORGANIC CHARGE-TRANSFER salts have been known for the wealth of electronic phases which result from the interplay of strong Coulomb correlations, significant electron-phonon interactions and low dimensionality [1], [2]. Prominent examples include Mott insulating, charge-ordered, local moment and itinerant magnetic states as well as superconductivity. Among the large variety of charge-transfer compounds which have been synthesized up to now, the salts (TMTTF)$_2$X, (TMTSF)$_2$X and (BEDT-TTF)$_2$X represent the most intensively studied systems. Here TMTTF, TMTSF and BEDT-TTF (or simply ET) stand for bis(tetramethyl)tetrathiafulvalene, bis(tetramethyl)tetraselenafulvalene, and (bis-(ethylenedithio)tetrathiafulvalene, respectively, and X for a monovalent anion. The κ-phase (BEDT-TTF)$_2$X salts constitute a particularly interesting class of materials due to their close proximity to the Mott metal-insulator transition and the presence of frustrating magnetic interactions [3]. In this paper we focus on the Mott insulator κ-(BEDT-TTF)$_2$Cu[N(CN)$_2$]Cl, cf. Fig. 1, which had been intensively studied in the past due to its model character for exploring Mott physics [4]-[6] and the pressure-induced superconductivity [7], [8], yielding the highest $T_c$ = 12.8 K within this class of materials. This system shows local moment antiferromagnetic ordering below $T_N$ ≈ 25-30 K, with spins aligned antiferromagnetically within the ET layers (*ac* planes) and ferromagnetically along the interlayer *b* axis [4]. At a temperature around 23 K a weak ferromagnetic canting was observed [4], [9]. In a recent work, Lunkenheimer *et al*.

[10] reported on another remarkable property of this material: at about the same temperature $T_N$ where spin order sets in, ferroelectricity was observed, making this system the first example of a multiferroic charge-transfer salt. In that work it was proposed that, in contrast to the frequently observed spin-driven multiferroics, where the onset of helical spin order triggers the formation of displacive ferroelectric order [11]-[14], a charge-order mechanism is at work here. In fact, a charge-order driven ferroelectricity has been proposed for charge-transfer salts [15] and experimentally observed in a variety of compounds such as in the (TMTTF)$_2$X series [16], [17]. Thus, as a possible scenario to account for the simultaneous occurrence of ferroelectricity and magnetic order in the present κ-(BEDT-TTF)$_2$Cu[N(CN)$_2$]Cl salt, it was suggested that charge order induces a collective off-centre positioning of the spins, thereby reducing the degree of frustration and enabling magnetic order to be stabilized [10]. So far, attempts to experimentally resolve the charge disproportionation have remained unsuccessful. According to infrared vibrational spectroscopy [18], any charge disproportionation, if present, should be less than ± 0.005 *e*, the resolution of their experiment. In [19] an alternative mechanism involving short-range discommensurations of the commensurate antiferromagnetic phase and charged domain-wall relaxations in the weak ferromagnetic state was proposed to account for the occurrence of an anomaly in the dielectric constant.

In this paper we report further experimental data which may help clearing up the above controversy. To this end we have performed measurements of various quantities including the dc conductivity, the magnetic susceptibility as well as the dielectric constant all on the same κ-(BEDT-TTF)$_2$Cu-[N(CN)$_2$]Cl single crystals whenever possible. We will address issues of the frequency dependence of the dielectric constant $\varepsilon'$ and the dielectric loss $\varepsilon''$ and the conclusions which can be drawn from these results with respect to the proposed relaxation processes [19]. In addition, we discuss sample-to-sample variations by comparing the results on various crystals prepared by using different electrochemical methods. The paper is organized as follows. After giving details of the experimental techniques applied and the electrocrystallization methods used in chapter II, we present our experimental results and a discussion in chapter III. The paper is concluded by a short summary in chapter IV.

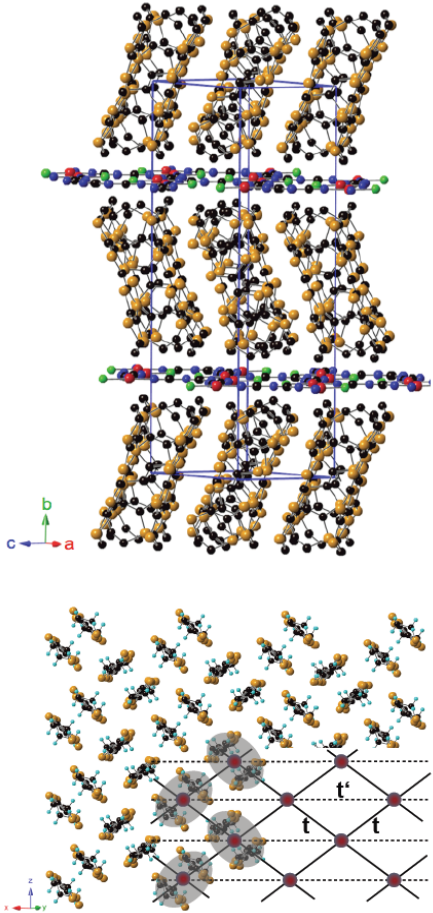

Fig. 1. Crystal structure of κ-(BEDT-TTF)$_2$Cu[N(CN)$_2$]Cl. Top: side view showing the layered structure consisting of thick layers of ET molecules separated by thin insulating anion layers. The unit cell is indicated by blue lines. In the structure, hydrogen atoms are omitted for clarity. Bottom: ET layers with the κ-type packing motif of the ET molecules viewed along their long axis. Selected dimers are highlighted by grey ellipses. In the effective-dimer model (low right corner) each dimer is represented by a single site (red spheres) on a distorted triangular lattice characterized by transfer integrals labeled $t$ and $t'$.

## II. EXPERIMENTAL

### 1) Crystal growth

Single crystals of κ-(BEDT-TTF)$_2$Cu[N(CN)$_2$]Cl were grown by electrochemical methods following two slightly different protocols. Whereas route (i) follows the procedure discussed in [8], [20], method (ii) differs in the following parameters: As a solvent a 90/20 mixture of THF and absolute ethanol with a little amount of water (0.7%) was used. The crystal growth was carried out in a 120 ml electrochemical cell consisting of 3 compartments separated by 2 glass filters and by using plate-like Platinum electrodes (10 × 15 mm$^2$). A constant voltage of 1.8 V (initially 1.3 V for one day) was applied, resulting in a current of 5-13 µA. The experiments ran over a period of 3-6 weeks at a constant temperature of 20°C. Here we present results of four different samples: crystal BZ1003G (corresponding to sample 1 in [10]), crystal mp1065, crystal MTB1063rod and crystal AF063I. While the former three crystals were prepared according to route (i), the latter one was synthesized according to route (ii).

### 2) Dielectric measurements

For the dielectric measurements, electrical contacts were applied at opposite sides of the crystal using either graphite paste or evaporated gold. Measurements were performed both with electrical field perpendicular or parallel to the *ac* planes. The dielectric constant and frequency-dependent conductivity were determined by using a frequency-response analyzer (Novocontrol α-Analyzer) and an autobalance bridge (Hewlett-Packard HP4284). With these techniques the conductivity measurements were conducted down to frequencies of $\nu$ = 2 Hz and 20 Hz, respectively. Since for frequencies below about 100 Hz no frequency dependence was observed, the conductivity at the lowest measured frequency was considered to represent the dc conductivity $\sigma_{dc}$.

### 3) Magnetic measurements

The measurements of the magnetic susceptibility $\chi$ were conducted by using a superconducting quantum interference device (SQUID) magnetometer (Quantum Design MPMS). The susceptibility data were corrected for the contribution of the sample holder.

## III. RESULTS AND DISCUSSION

### A. The proof of ferroelectricity in κ-(BEDT-TTF)$_2$Cu-[N(CN)$_2$]Cl

In this paragraph, we briefly summarize the experimental results presented in [10] which prove the transition into ferroelectric order in κ-(BEDT-TTF)$_2$Cu[N(CN)$_2$]Cl. Three pieces of evidence were provided. (i) First, a pronounced peak was found in the temperature dependence of the dielectric constant $\varepsilon'$ for all three crystals studied in [10]. This is shown in Fig. 2 exemplarily for crystal BZ1003G at various frequencies ranging from 1 MHz down to 2.1 Hz.

In these measurements the electric field $E$ was applied perpendicular to the ET layers. Whereas the peak height, reaching values of several hundred at low frequencies, decreases upon increasing the frequency, the peak position remains almost unaffected with varying the frequency. This frequency response is characteristic of order-disorder-type ferroelectrics, where upon cooling through a phase transition temperature $T_{FE}$ electric dipoles, which are disordered at high temperatures, order with a net overall polarization [21]. The dashed line in Fig. 2 corresponds to a Curie-Weiss fit, $\varepsilon_s = C/(T - T_{CW}) + \varepsilon_b$, with a Curie-Weiss temperature $T_{CW} \approx 25$ K and a background dielectric constant $\varepsilon_b$. The high value for $\varepsilon_b$ of about 120 is predominantly due to stray-capacitance contributions resulting from the small sample dimensions. A qualitatively similar behavior, with a slightly more rounded peak, was obtained for electric field aligned parallel to the $ac$ planes. The rounding observed for this configuration was attributed to the strongly anisotropic conductivity of the material, with in-pane conductivities exceeding the out-of-plane values by two to three orders of magnitude, making in-plane measurements difficult. (ii) The second piece of evidence was derived from so-called positive-up-negative-down (PUND) measurements [10]. In these runs, performed at 25 K, an electric field which increases linearly in time was applied to the sample while measuring the time-dependent current across the sample. At maximum field amplitude $|E|_{max}$, $E$ was kept constant for some time and then ramped down at the same rate. Then, after a delay of 0.2 s, a second identical trapezoid pulse was applied to the sample. It was found that the resulting current response shows a sharp peak only for the first pulse once the electric field exceeds a threshold value of $|E| \approx 10$ kV/cm. In the second pulse no such peak was observed. However, in a consecutive third experiment, where the polarization of the electric field was reversed, the peak could be reproduced while there was no peak in the fourth run keeping the same polarization. This set of experiments demonstrates that by the application of a sufficiently strong electric field the macroscopic polarization of the sample becomes switched, i.e., the electric dipoles within the ferroelectric domains are forced to align with the field. Since this implies a motion of charges, a peak in the current results. Once the domains are oriented, as is the case after the first and third experiments, the field sweep with the same field polarization will not cause any reorientation of the domains and thus will not cause any peak in the current. (iii) In addition, field-dependent polarization measurements $P(E)$ were carried out. For measurements performed at temperatures above the ferroelectric transition, elliptical curves were observed, typical for paraelectricity with some loss contribution due to charge transport. However, at temperatures slightly below $T_{FE}$, a clear onset of nonlinearity above about 8.5 kV/cm and saturation at highest fields with a saturation polarization of about 0.4 μCcm$^{-2}$ was observed [10].

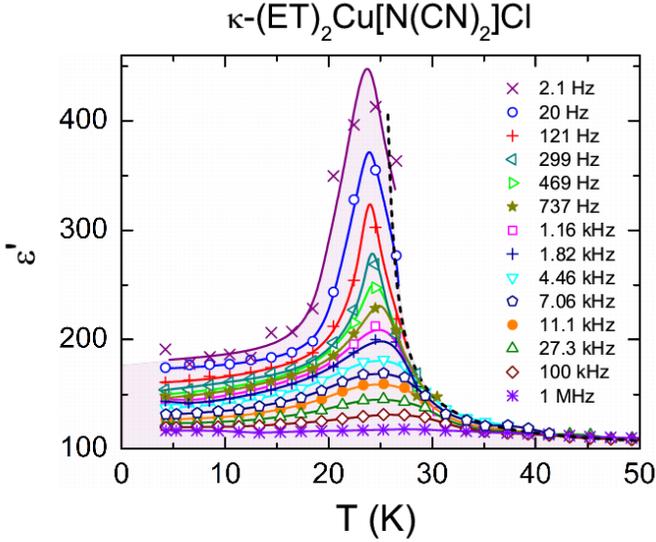

Fig. 2. Temperature dependence of the dielectric constant $\varepsilon'$ of κ-(BEDT-TTF)$_2$Cu[N(CN)$_2$]Cl crystal BZ1003G (sample 1 in ref. [10]) for selected frequencies measured with electrical field $E$ parallel to the out-of-plane $b$-axis. Compared to Fig. 2 of [10], additional frequencies are shown. Broken line represents a Curie-Weiss fit.

Moreover, an important conclusion as for the nature of the ferroelectric order could be drawn from the behavior of the dielectric anomaly in a magnetic field. It was found that by increasing the magnetic field from 0 to 9 T, i.e., from fields below to distinctly above the spin-flop field of $B_{sf} = 0.25$ T, the dielectric constant remained entirely unaffected [10]. From this result, a spin-driven mechanism, revealed in various multiferroics with helical spin order, could be ruled out: in these systems the spin-driven polarization can be ascribed to the electromagnetic coupling $P \propto (S_i \times S_j) \times Q$ with $Q$ denoting the propagation vector of the spin order and $S_i$, $S_j$ adjacent spins [11], [12]. Thus, from the absence of any effect on $P$ on crossing the spin-flop transition, which drastically changes the cross product $S_i \times S_j$, rules out that a spin-driven mechanism is at work for the present material [10]. The above experiments provide strong evidence for a ferroelectric transition in κ-(BEDT-TTF)$_2$Cu[N(CN)$_2$]Cl. At the same time, the fact that $T_{FE}$ virtually coincides with the antiferromagnetic transition strongly suggests that both phenomena are intimately coupled to each other.

### B. Frequency dependence of the dielectric permittivity

In [10] (and with additional frequencies in Fig. 2), the temperature dependence of the dielectric constant $\varepsilon'$ of κ-(BEDT-TTF)$_2$Cu[N(CN)$_2$]Cl is presented for various frequencies, providing clear evidence for order-disorder-type ferroelectricity. In Fig. 3, we show the frequency dependence of $\varepsilon'$ as obtained from the same experiments as the data shown in [10]. The figure shows results for the electrical field oriented perpendicular (Fig. 3(a)) and parallel (Fig. 3(b)) to the ET layers. For each configuration, two curves are presented, one measured at a temperature close to the peak in $\varepsilon'(T)$ and another one at a distinctly lower temperature. In all cases, $\varepsilon'(\nu)$ continuously increases with decreasing frequency. Obviously, there is no indication for saturation at low frequencies, and thus, no evidence for a relaxational process suggested in [22] and [19].

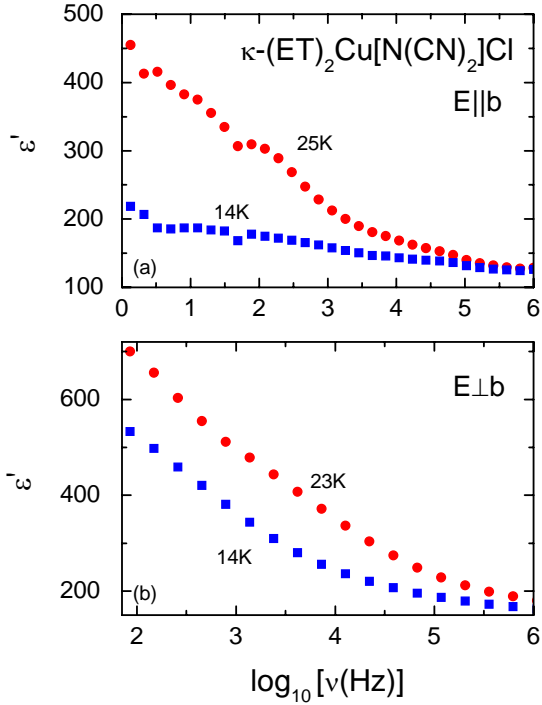

Fig. 3. Frequency dependence of the dielectric constant of κ-(ET)$_2$Cu[N(CN)$_2$]Cl crystal BZ1003G (sample 1 in [10]) as measured at two temperatures close to and below the $\varepsilon'(T)$ peak and for electrical fields oriented along the out-of-plane $b$ axis (a) and perpendicular to it (b), i.e., parallel to the ET layers. With decreasing frequency, $\varepsilon'(\nu)$ continuously increases without any indication for saturation at low frequencies. Note that saturation would be expected for a relaxational process.

As an example for the frequency dependence of the dielectric loss $\varepsilon''$, we show in Fig. 4(a) data for the electric field directed perpendicular to the ET layers. The data were taken at 25 K where the peak in $\varepsilon'(T)$ was observed [10]. Qualitatively similar behavior was also found for temperatures below or above the peak position. Figure 4(a) reveals a strong and continuous decrease of $\varepsilon''(\nu)$ with increasing frequency. Below about 10 kHz, a linear behavior with slope -1 is found in the double-logarithmic representation of Fig. 4(a), corresponding to a frequency dependence $\varepsilon'' \propto 1/\nu$. According to the general relation $\sigma' = 2\pi\nu\varepsilon''\varepsilon_0$ (with $\varepsilon_0$ the permittivity of vacuum), the observed $\varepsilon'' \propto 1/\nu$ behavior implies $\sigma' = $ const. Indeed the plot of $\sigma'(\nu)$, provided in Fig. 4(b), reveals a frequency-independent behavior at low frequencies. The conductivity obtained in this regime can be identified with the dc conductivity. Thus, a strong contribution from dc charge transport dominates the $\varepsilon''$ spectrum over wide ranges in frequency leading to giant loss values up to the order of 10$^4$. The temperature dependence of the dc conductivity was discussed in detail in [10].

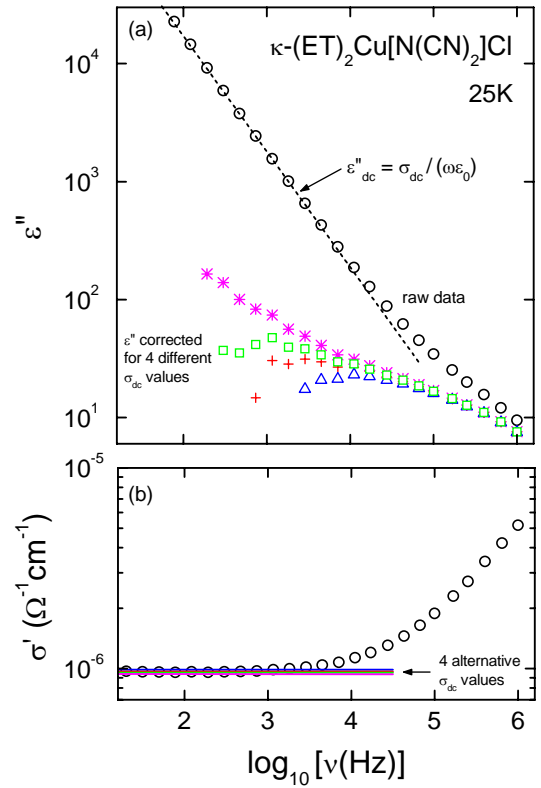

Fig. 4. Frequency dependence of the dielectric loss $\varepsilon''$ (a) and the conductivity $\sigma'$ of κ-(ET)$_2$Cu[N(CN)$_2$]Cl crystal BZ1003G (sample 1 in [10]) at 25 K and for $E\|b$. The open circles in (a) indicate raw data, directly related to $\sigma'(\nu)$, shown in (b), via the relation $\varepsilon'' = \sigma'/(2\pi\nu\varepsilon_0)$. The other curves (colored symbols) in (a) have been obtained by correcting for the dc conductivity via $\varepsilon''_{corr} = \varepsilon'' - \sigma_{dc}/(2\pi\nu\varepsilon_0)$. Four different values of $\sigma_{dc}$ were used which vary only slightly by 5% at maximum. These four choices of $\sigma_{dc}$ are indicated by the nearly indiscernible horizontal lines in (b). Obviously, the questions of whether or not a peak exists in the dielectric loss at all, and at what frequency such a peak occurs, strongly depend on the subtracted value of $\sigma_{dc}$.

At frequencies above about 10 kHz in Figs. 4(a) and (b) deviations from dc behavior are observed. These deviations can be described by another power law $\sigma'(\nu) = \sigma_{dc} + \sigma_0 \nu^s$ with exponents $s < 1$, and correspondingly, an exponent $s-1$ in $\varepsilon''(\nu)$. As already discussed in [10] (Supplementary Information), such power law behavior is typical for hopping conductivity [23], [24]. However, it should be noted that the high-frequency flanks of the loss peaks arising from dielectric relaxation processes, usually follow power laws as well [25], [26]. Therefore one might speculate about the presence of a relaxation peak in the spectrum of Fig. 4(a), whose left flank may be superimposed by the strong dc conductivity contribution. In such cases, $\varepsilon''$ data often are corrected for the dc contribution using the relation $\varepsilon''_{corr} = \varepsilon'' - \sigma_{dc}/(2\pi\nu\varepsilon_0)$ to reveal the "true", purely relaxational part of the dielectric loss. Indeed, such a correction was performed in [22] and [19] for dielectric data on κ-(BEDT-TTF)$_2$Cu[N(CN)$_2$]Cl and the presence of loss peaks was claimed for various temperatures. In [22], the obtained corrected data were interpreted to provide evidence for two relaxational processes. In contrast, in [19] the spectra were fitted assuming a single relaxation process only.

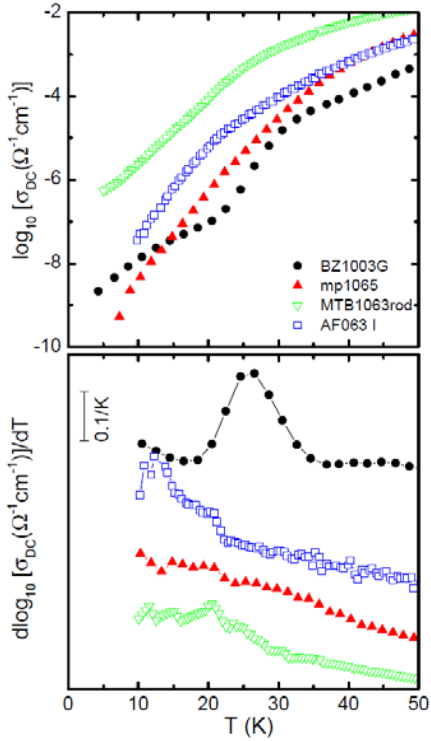

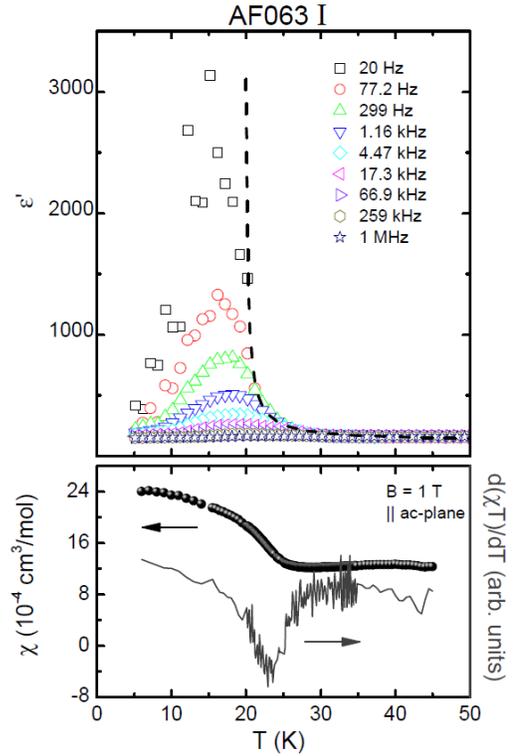

Fig. 5. Top panel: dc conductivity vs temperature in a semilogarithmic representation for four different single crystals with $E$ along the out-of-plane $b$ axis. Lower panel: temperature derivative of $\log\sigma_{dc}$ in the same temperature range as shown in the top panel. The data sets have been shifted vertically for clarity.

Fig. 6. Top panel: dielectric constant $\varepsilon'$ at selected frequencies; dashed lines corresponds to a Curie-Weiss behavior with $T_{CW}$ = 20 K. Bottom panel: magnetic susceptibility $\chi$ (left axis) and derivative of $\chi \cdot T$ (right axis) on the same $\kappa$-(BEDT-TTF)$_2$Cu[N(CN)$_2$]Cl crystal (AF063 I). The susceptibility data were taken in a magnetic field of 1 T applied parallel to the $ac$-plane.

This process was ascribed to "charged domain-wall relaxations in the weak ferromagnetic state at lower temperatures".

In Fig. 4(a), we show the results of a dc correction procedure applied to our data at 25 K (at other temperatures, the outcome is similar). Generally, when correcting loss spectra for charge-transport contributions, the proper choice of the absolute values of the dc conductivity represents a major problem. In Fig. 4(b), the horizontal lines indicate four possible choices for $\sigma_{dc}$. These lines, corresponding to $\sigma_{dc}$ values varying only by about 5% at maximum – a variation which roughly corresponds to the experimental error - are nearly indiscernible. Nevertheless, the corrected loss spectra shown by the four lower curves in Fig. 4(a) (colored symbols) are markedly different. Obviously, in $\kappa$-(BEDT-TTF)$_2$Cu-[N(CN)$_2$]Cl the questions whether or not a peak exists in the dielectric loss and what the peak position might be, strongly depend on the particular choice of $\sigma_{dc}$. Therefore statements referring to such a peak are of limited significance only. According to our experience, these problems always arise for materials with relatively high dc conductivity, where the corrected data are of much lower magnitude than the dc-dominated raw data (cf. Fig. 4(a)). In summary, when taking into account the absence of saturation in $\varepsilon'(\nu)$ at low frequencies (Fig. 3) and the ambiguities in the correction of $\varepsilon''(\nu)$ for the strong dc conductivity (Fig. 4), we note that there is no evidence for a relaxational process in $\kappa$-(BEDT-TTF)$_2$Cu[N(CN)$_2$]Cl.

### C. Sample-to-sample variations

In Fig. 5 we show a compilation of low-temperature conductivity results, in a semi-logarithmic representation, measured along the out-of-plane $b$ axis for the different $\kappa$-(BEDT-TTF)$_2$Cu[N(CN)$_2$]Cl single crystals studied in this work. At higher temperatures (not shown) the data for all crystals reveal the well-known semiconducting-like behavior [7], [8], [27] characterized by a gradual decrease upon cooling from 300 K. This gradual loss of conductivity upon cooling is accelerated below about 50 K due to an increase of the charge gap. The particular focus of the present study lies on the temperature range from 20 - 30 K where an abrupt decrease in the dc conductivity by about two orders of magnitude was found in [10] for crystal BZ1003G, cf. solid black circles in Fig. 5(a). This feature manifests itself in a slightly broadened maximum in the derivative $d(\log\sigma_{dc})/dT$, shown in Fig. 5(b). Although a peak in $d(\log\sigma_{dc})/dT$ is not observed in the data for the other crystals, a more or less pronounced increase in $d(\log\sigma_{dc})/dT$ at a temperature slightly above 20 K can be discerned in all data sets. Thus the results of Fig. 5 confirm the presence of an anomaly in the dc conductivity for $\kappa$-(BEDT-

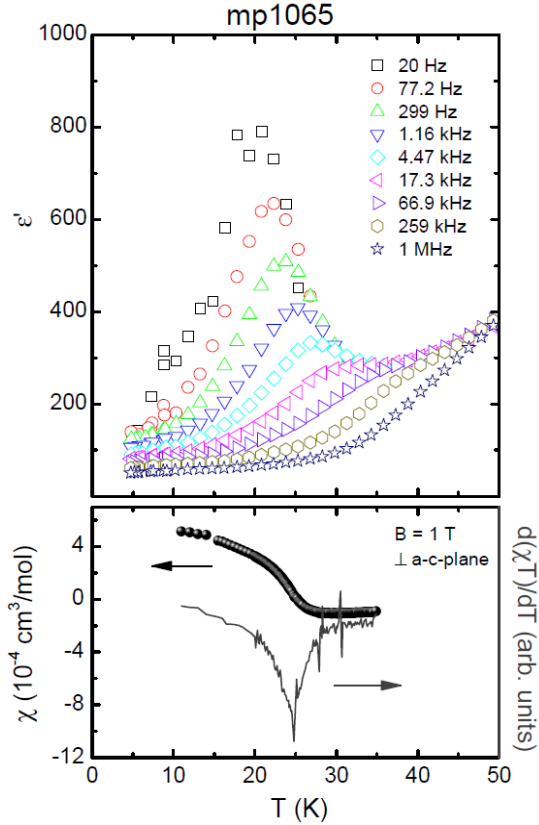

Fig. 7. Results of the dielectric constant $\varepsilon'$ at selected frequencies (top panel) and magnetic susceptibility $\chi$ (left axis) and derivative of $\chi \cdot T$ (right axis) on the same κ-(BEDT-TTF)$_2$Cu[N(CN)$_2$]Cl crystal (mp1065). The susceptibility data were taken in a magnetic field of 1 T applied perpendicular to the ac-plane.

TTF)$_2$Cu[N(CN)$_2$]Cl at low temperature which manifests itself either in a sudden drop of $\sigma_{dc}$ or a more rapid decrease of $\sigma_{dc}$ upon cooling. The data also show that there is a significant sample-to-sample variation with regard to the position, shape and size of the effect.

In Figures 6 and 7 we show the results of the dielectric constant (upper panels) together with the magnetic susceptibility (lower panels) for the crystals AF063I and mp1065, respectively, for temperatures $T \leq 50$ K. For both crystals we find a pronounced peak in the dielectric constant with a peak height at the lowest frequency of 20 Hz in excess of the one found for crystal BZ1003G. With increasing frequency the peak height in both cases becomes drastically reduced. As for the position of the peaks, however, two distinctly different behaviors were observed. For the crystal AF063I, the peak position is almost unaffected by frequency, similar to the observation made in [10] for all three crystals studied there including crystal BZ1003G. In contrast for crystal mp1065 the peak position increases with increasing frequency. Qualitatively similar, frequency-dependent behavior was observed for two other crystals (not shown) taken from the same batch. Overall, the temperature and frequency dependence of $\varepsilon'$ in these samples is typical for so-called relaxor ferroelectricity, which often has been interpreted in terms of short-range cluster-like ferroelectric order [28], [29]. At the same time, the magnetic susceptibility for both crystals shown in Figs. 6 and 7 looks rather similar. The data reveal a slightly broadened increase below about 25 K, consistent with literature results [4], indicating the transition into a canted antiferromagnetic state. The lower panels of the figures also include the derivative of the product $\chi(T) \cdot T$ from which the magnetic ordering temperature $T_N$ can be estimated to 25 K (crystal mp1065) and 23 K (AF063I).

The above results on the dc conductivity, the dielectric constant and the magnetic susceptibility on various single crystals of κ-(BEDT-TTF)$_2$Cu[N(CN)$_2$]Cl prove beyond doubt (i) the existence of a pronounced dielectric anomaly at a temperature very close to the magnetic ordering temperature. (ii) In five out of eight crystals studied here and in [10] via dielectric measurements, this dielectric anomaly is found to be of order-disorder type, characterized by a frequency-independent peak in $\varepsilon'$, reflecting a transition into a long-range ordered ferroelectric state. This contrasts with a relaxor-type of ferroelectricity, indicative of short-range ferroelectric correlations, revealed for three crystals including crystal mp1065. (iii) Measurements of the dc conductivity on those crystals yielding an order-disorder-type ferroelectricity reveal a sudden change (drop or break in the slope) at a temperature close to the position of the maximum in $\varepsilon'$. In contrast, the anomaly in the dc conductivity for the crystal showing the relaxor-type ferroelectricity is less clearly pronounced, if present at all. For the two crystals AF063I and mp1065, representing respectively order-disorder- and relaxor-type ferroelectricity, the signatures in the magnetic susceptibility are rather similar, yielding a rapid increase below about 23-25 K, indicative of long-range magnetic order with ordering temperatures differing by about 2 K. The so-derived Néel temperatures appear to be slightly higher than the position of the peak in $\varepsilon'(T)$. For the crystal AF063I, showing order-disorder ferroelectricity, the ferroelectric transition temperature $T_{FE}$ can be estimated from a fit of the high-temperature flanks of the $\varepsilon'(T)$ peaks at low frequencies (i.e., the static dielectric constant $\varepsilon_s(T)$) using a Curie-Weiss law, $\varepsilon_s = C/(T-T_{CW}) + \varepsilon_b$ (broken line in Fig. 6; $\varepsilon_b$ is an additional background term, see [10]). This leads to a Curie-Weiss temperature $T_{CW} = 20$ K $\approx T_{FE}$. Thus, here $T_{FE}$ and $T_N \approx 23$ K are of comparable magnitude. For relaxor ferroelectrics (sample mp1065), showing a diffuse phase transition only [28], the unequivocal determination of a transition temperature is difficult. In any case, Fig. 7 is consistent with the reasonable assumption that the short-range ferroelectric order occurring in the relaxor state is sufficient to break the magnetic frustration, leading to long-range magnetic order. Overall, the present data disclose significant sample-to-sample variations for this material, which seem to affect the dielectric properties and the conductivity more than the magnetic properties.

We stress that the order-disorder type ferroelectricity revealed here for the majority of crystals investigated is clearly incompatible with the scenario put forward in [19] where charged domain walls were made responsible for the dielectric signatures observed in their study. In light of the above-mentioned sample-to-sample variation, however, it is entirely possible that the crystal studied there showed a relaxor-type dielectric response which led the authors to consider an inhomogeneous, short-ranged-ordered ferroelectric state.

The present results, i.e., the verification of an order-disorder type of ferroelectricity in κ-(BEDT-TTF)$_2$Cu[N(CN)$_2$]Cl accompanied by a sudden change of the dc conductivity, are consistent with the scenario suggested in [10]. There it was proposed that charge order induces a collective off-centre positioning of the spins, thereby reducing the degree of frustration which triggers the magnetic order. One may conjecture that the lack of direct experimental evidence for the predicted charge order is due to the small size $\pm \delta\rho$ of the accompanying charge disproportionation within the ET dimers. A rough estimate of $\delta\rho$ can be obtained by comparing the size of the dielectric anomaly observed in the present study and in [10], where $\varepsilon'$ reaches values up to 3000 at lowest frequencies, with the peak heights revealed for the ferroelectric transition in the related quasi-onedimensional (TMTTF)$_2$X salts. For the latter, $\varepsilon'$ values of about $2\cdot10^6$ for X = AsF$_6$ and about $8\cdot10^5$ for X = PF$_6$ were observed [17]. From infrared conductivity measurements for these salts a charge disproportionation $\delta\rho$ of $\pm$ 0.13 $e$ (X = AsF$_6$) and $\pm$ 0.06 $e$ (PF$_6$) was inferred [30]. By assuming that $\delta\rho$ scales linearly with the size of the dielectric anomaly, a charge disproportionation $\delta\rho$ distinctly smaller than 0.005 $e$ can be expected for the present material. This is consistent with the experimental results reported in [18] where a charge disproportionation $\delta\rho$ in excess of $\pm$ 0.005 $e$ was ruled out.

IV. SUMMARY

By a comparative study covering different single crystals of κ-(BEDT-TTF)$_2$Cu[N(CN)$_2$]Cl we have provided further evidence in support of an order-disorder type ferroelectricity in this material. The ferroelectric order, accompanied by a sudden change in the dc conductivity, is found to be located very close to the transition to long-range antiferromagnetic order, indicating a type of multiferroicity where both orders are intimately coupled. The data are consistent with the scenario of a charge-order-driven magnetism proposed in [10] but clearly incompatible with the inhomogeneous state considered in [19] where the dielectric response were ascribed to a relaxation of charged domain walls. By comparing the size of the dielectric anomaly observed in the present study with those of well-established charge-order-driven ferroelectrics, a charge disproportionation of $\pm\delta\rho$ distinctly smaller than 0.005 $e$ is expected, consistent with the estimates from recent optical experiments [18]. At the same time our study reveals the existence of a considerable sample-to-sample variation in the conductivity and dielectric properties with some crystals showing the characteristics of relaxor-type ferroelectrics. It is left to further studies to unravel the detailed interplay of charge and spin degrees of freedom in this material. In particular, it has to be shown whether a small charge disproportionation is sufficient to trigger a long-range antiferromagnetic state on the frustrated triangular lattice.

ACKNOWLEDGMENT

This work was supported by the Deutsche Forschungsgemeinschaft through the Transregional Collaborative Research Centers TRR 49 and TRR 80. Work at Argonne was supported by the US Department of Energy Office of Science, operated under contract no. DE-AC02-06CH11357